\documentclass[proof]{WileyASNA-v1}
\UseRawInputEncoding
\articletype{Article Type}%

\received{* * 2022}
\revised{* * 2022}
\accepted{* * 2022}

\begin{document}

\title{Extreme Accretion Events: TDEs and Changing-Look AGN}

%% Dear editors, please do not spell out authors' first names. 
\author[1]{S. Komossa*}
\author[2]{D. Grupe}

\authormark{S. KOMOSSA \& D. GRUPE }

\address[1]{\orgdiv{Max-Planck-Institut f\"ur Radioastronomie}, 
%% \orgname{Org name}, 
\orgaddress{\country{Bonn, Germany}}}
\address[2]{\orgdiv{Deptartment of Physics, Geology, and Engineering Technology, Northern Kentucky University, Kentucky, USA}} 
%\orgname{Org name}, \orgaddress{\state{State name}, \country{Country name}}}

%\address[1]{\orgdiv{Max-Planck-Institut f\"ur Radioastronomie}, 
%% \orgname{Org name}, 
%\orgaddress{\state{Auf dem H\"ugel 69}, \country{53111 Bonn, Germany}}}
%\address[2]{\orgdiv{Department of Physics, Geology, and Engineering Technology, Northern Kentucky University, 1 Nunn Dr, Highland Heights, KY 41099, USA}} 

\corres{*S. Komossa. \email{astrokomossa@gmx.de}}

\abstract{We present a review of the topics of X-ray stellar tidal disruption events (TDEs) and changing-look active galactic nuclei (AGN).  
Stars approaching a supermassive black hole (SMBH) can be tidally disrupted and accreted. TDEs were
first discovered in the X-ray regime and appear as luminous, giant-amplitude flares from inactive galaxies. The early X-ray observations with ROSAT also established the extreme X-ray spectral softness of these events with temperatures of order 50-100 eV that continues to be seen in the majority of more recently identified events.  
While the majority of X-ray TDEs
has been identified from {\it inactive} galaxies and some showed the highest
amplitudes of variability recorded from galaxy cores (amplitudes exceeding factors of 1000--6000), a small fraction of {\it active}  galactic nuclei (AGN) has been found
to be highly variable as well.
%(amplitudes exceeding factors of 100). 
In AGN, this so-called
changing-look phenomenon often comes with a strong change in the optical broad
emission lines, leading to Seyfert-type changes between class 1 and class 2.
These two forms of activity represent the extremes of variability among active and quiescent galaxies, and have opened up a new window on understanding accretion physics under extreme conditions.
Finally, we introduce the term ``frozen-look AGN'' to describe systems that show constant line emission despite strong/dramatic changes in the observed ionizing continuum. These systems are best explained by strong changes of absorption along our line-of-sight.   
% Here we provide a review of the field, with emphasis on the important XMM-Newton contributions to this topic.
}


\jnlcitation{\cname{%
\author{Komossa S.}, 
\author{D. Grupe}} (\cyear{2022}), 
\ctitle{Extreme Accretion Events: TDEs and Changing-Look AGN}, \cjournal{Astronomische Nachrichten (invited review, proceedings of the 2022 XMM-Newton science workshop on ``Black Hole Accretion'')}, \cvol{2022;00:0--0}.}

%%\fundingInfo{Funding info text.}

\maketitle

\section{Introduction}\label{sec1}

%Many authors submitting to NJD (New Journal Design) journals use \LaTeXe\ to
%prepare their papers. This paper describes the
%\textsf{WileyASNA-v1.cls} class file which can be used to convert articles produced with other \LaTeXe\ class files into the correct form for publication in \emph{Wiley NJD Journals}.

Both, stellar tidal disruption events (TDEs) and changing-look active galactic nuclei (AGN) represent extreme forms of accretion onto supermassive black holes (SMBHs). An important main distinction is the 
% different host galaxies.
absence (quiescent host galaxies) or presence (AGN) of a long-lived accretion disk. 
TDEs have been preferentially identified in quiescent hosts, with dormant SMBHs. A long-lived accretion disk is absent, and therefore long-lived optical broad emission lines from the broad-line region (BLR) and high-ionization narrow emission lines from the narrow-line region (NLR) are absent as well.  
In contrast, AGN harbor a long-lived accretion disk, and show characteristic broad and narrow optical emission lines. The BLR near the SMBH follows changes in the ionizing continuum. The emission lines from the NLR
are insensitive to rapid continuum flaring because of the large distance, large spatial extent and low gas densities (and correspondingly long recombination timescales on the order of 1000s of years). The NLR is therefore an ideal tracer of long-lived accretion activity.  

While TDEs can happen in both, quiescent galaxies and AGN, it has been convincingly argued that TDEs are most reliably identified if they are detected from {\it quiescent} galaxies \citep[e.g.][]{Rees1989}, since variability due to a long-lived accretion disk can then be safely excluded, and thus this approach has been followed since the discovery of the first TDEs in X-rays \citep{KomossaBade1999}. A first important step in the identification of any new TDE has therefore been the acquisition of optical spectra to exclude the presence long-lived AGN activity. 
{\footnote{More recently, this criterion has been somewhat relaxed, and occasionally TDE candidates have also been reported in AGN.}}. 

With this important distinction in mind, we first provide a review of TDEs from quiescent galaxies, and then review the changing-look phenomenon in highly variable AGN. 

\begin{figure}
\centering
\includegraphics[clip, width=\columnwidth]{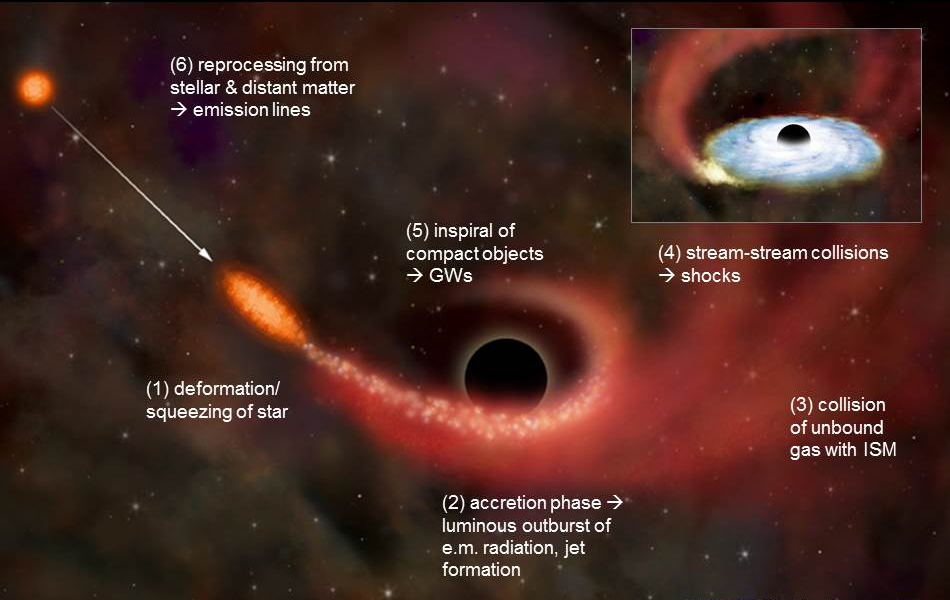}
    \caption{Physical processes and emission mechanisms during the main stages of stellar tidal disruption, overlaid on an artist's image. Image credit: NASA/CXC/M. Weiss/ \citet{Komossa2004}. 
    }
    \label{fig:TDE-artist}
\end{figure}

\section{Stellar tidal disruption events}

\subsection{Theory and first detections}

Stars can be tidally disrupted by SMBHs when the tidal forces of the black hole exceed the self gravity of the star. The subsequent accretion of the stellar material then causes a bright flare of electromagnetic radiation with a peak luminosity that can reach the Eddington luminosity given the high initial gas supply rate \citep[][our Fig. \ref{fig:TDE-artist}]{Rees1988}.   

TDEs were first predicted based on theoretical considerations in the 1970s. 
Important aspects of the analytical theory of TDEs and their evolution were formulated around that time \citep[e.g.][]{Hills1975}, and first speculations emerged where TDEs might play a role in astrophysical environments, for instance as contributors to SMBH growth 
\citep{FrankRees1976} or as an explanation for certain types of gamma-ray bursts \citep{CarterLuminet1982}.   

Main aspects of TDE theory (of solar-type stars) were developed well before the first events were identified in observations. 
After disruption, the stellar material is spread over
a range of Keplerian orbits of high excentricities with a return rate to pericenter that scales as d$M$/d$t$ $\propto t^{-5/3}$.  
Due to stream-stream collisions 
(effectively high viscosity regime) 
the material quickly circularizes, and if accreted as quickly as it returns, then the observed X-ray light curve declines as $L \propto t^{-5/3}$ \citep{Rees1990}. 
% The mass-return rate can exceed the Eddington. 
Only a small fraction of the stellar material (much less than 50\%) is expected to be ultimately accreted. About 50\% of a solar-mass star is initially unbound, and more material becomes unbound as the stellar debris evolves \citep{Ayal2000} through stream-stream collisions and through radiation-pressure driven disk winds.     

Because of the different dependence of tidal radius and Schwarzschild radius on SMBH mass, solar-type stars are no longer disrupted for SMBH masses larger than $\sim$${10^8}$ M$_{\odot}$; instead they are swallowed whole. The exact value also depends on SMBH spin and stellar orbit \citep{Beloborodov1992}.  Disruption of white dwarfs requires $M_{\rm BH} \le 10^5$ M$_{\odot}$. 

At 10$^{6-7}$ M$_{\odot}$, an Eddington-limited event would produce a luminous flare,
easily detectable in the nearby universe. Besides brightness, the other property relevant for detection of events is the TDE rate. In addition to SMBH mass and spin, rates depend on the types of stars, and stellar orbits and density in the center of the host galaxy. Rates are on the order of $10^{-4}-10^{-5}$/yr events per galaxy \citep[e.g.][]{Rees1988, WangMerritt2004}. Experiments with large-area sky coverage are therefore ideal in identifying TDEs. 
%% or deep coverage of dense fields, or very deep fields. 

\begin{figure}
\centering
\includegraphics[clip, width=\columnwidth]{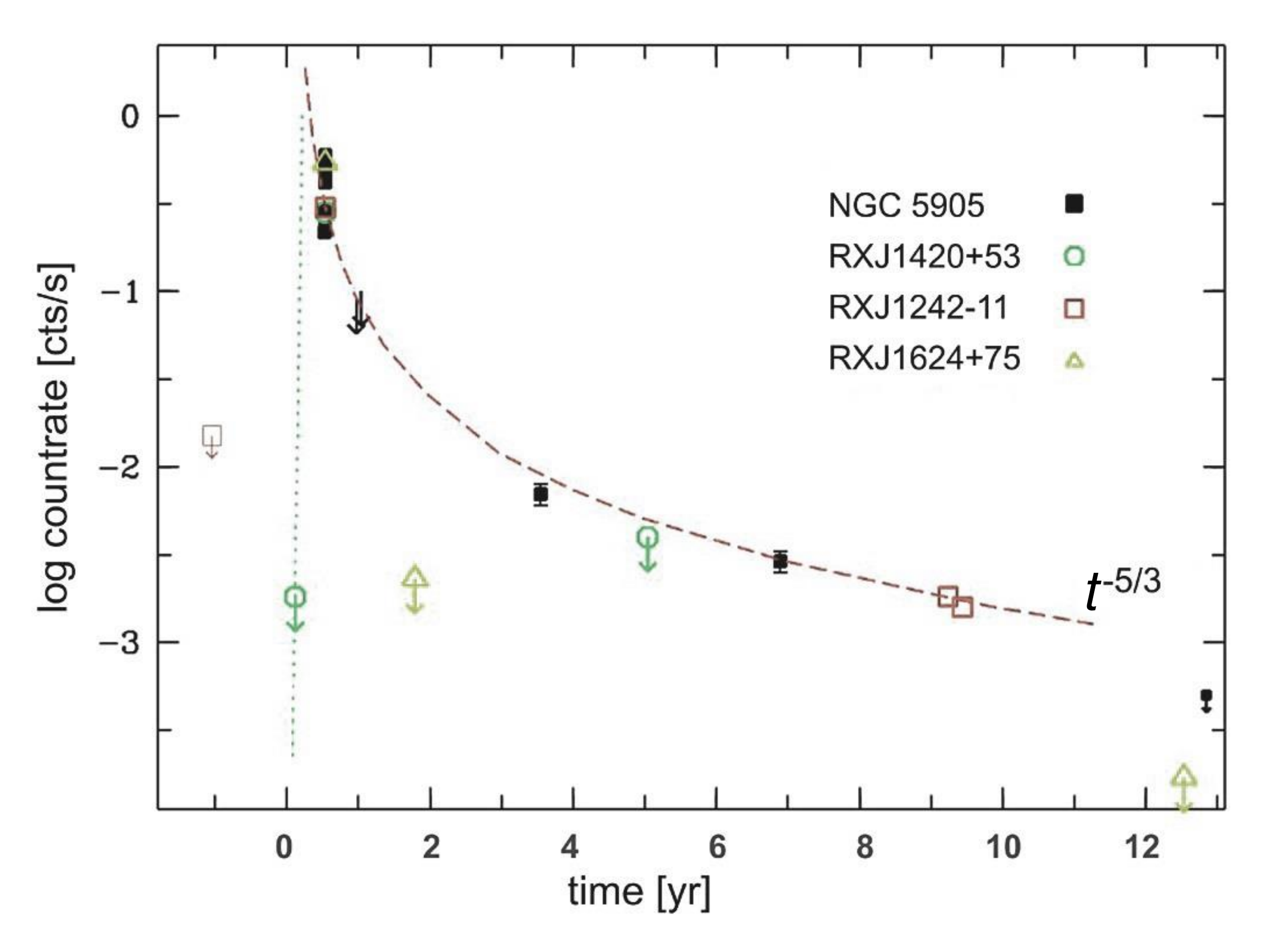}
    \caption{X-ray light curve of four TDEs based on ROSAT, XMM-Newton and Chandra observations. All events were shifted to the same peak time. }
    \label{fig:ROSAT}
\end{figure}

Such an experiment was the soft (0.1--2.4 keV) X-ray all-sky survey carried out with the X-ray mission ROSAT \citep{Truemper1982}. Four TDEs were initially discovered \citep[][our Fig. \ref{fig:ROSAT}]{Bade1996, KomossaBade1999, KomossaGreiner1999, Grupe1999, Greiner2000}, plus a few more in the ROSAT archive after the end of the mission \citep[e.g.][]{Cappelluti2009, Maksym2014, Hampel2022}. These events had the properties predicted much earlier by theory: huge X-ray peak luminosities reaching at least $\sim 10^{43-44}$ erg/s in the soft X-ray band alone, peaking in the EUV--soft X-rays, with declines on the order of months to years consistent with $L \propto t^{-5/3}$, and quiescent host galaxies. 
% 
% The early TDE discoveries firmly established the super-soft X-ray spectra of the events at peak luminosity with temperatures around 100 eV  \citep{Komossa2002}, more recently observed in the large majority of newly identified X-ray TDEs as well (Sect. 2.3).   

Detailed predictions of the spectral shape of TDEs were not yet available from theory. 
The first X-ray TDE discoveries firmly established the extreme X-ray spectral softness near peak of 40--100 eV \citep[highlighted particularly  by][]{Komossa2002}  
that has more recently been seen in most newly-identified X-ray events and that is often used as an additional TDE search criterion besides amplitude of variability. 
%% and that has influenced the choice of soft x-ray band range in future missions. 

Initially one main motivation was to use the luminous, transient accretion flares of TDEs to identify dormant SMBHs at the centers of distant, quiescent galaxies \citep{Rees1988}, since such SMBHs would not reveal their presence in any other easily detectable way. 
Meanwhile, TDE statistics and the physics in their different regimes of evolution (Fig. \ref{fig:TDE-artist}), allow us to address many more questions 
of broad relevance for 
extragalactic astrophysics (Sect. 2.2). These are 
aided by simulations 
that have not only allowed to confirm some of the assumptions in the early analytical models 
but have also allowed to expand the regime of TDE parameter space to include different types of disrupted stars, different stellar orbits including high penetration factors, extreme stellar compression, spinning and non-spinning SMBHs, single and binary stars, and single and binary SMBHs;  some regimes only accessible numerically in very advanced computer simulations \citep[][and references therein]{Andalman2022, BonnerotWenbin2022, Bu2022, CoughlinNixon2022, Cufari2022, Ryu2022b, Wen2022} with many new predictions of observable facets of TDEs.  
% Bu, De-Fu 2022: super-Edd accretion to explain opt-UV
% Ryu: relativistic events. basically predicts the X-ray events already observed.
% Wen s., 2022: modelling with emphasis on X-spectra. 
% Andalman2022: strems-strems collisions, disk formation, stream disruption to explain flaring in Swift1644-fashion. 
% CoughlinNixon2022: strong penetration and stellar compression
% These go along with advances in X-ray detection methods and instrument sensitivity.   

\subsection{New ongoing and future applications of TDE observations} 

The theoretical advances along with advances in X-ray detection methods and instrument sensitivity, allow us to address important new questions and applications related to TDEs. A few of them are highlighted below (see \citet{Komossa2012} for a longer list).

\paragraph{BH spin}
TDE rates and light curve shapes depend on SMBH spin which can therefore be estimated once densely covered light curves and good TDE statistics are available \citep{Kesden2012}. 

\paragraph{Binary SMBH identification and orbital-parameter measurements}
Binary SMBHs characteristically change TDE light curves and these can therefore be used to identify binaries and measure their orbital parameters.
Since the secondary SMBH perturbs the stream of stellar material, the accretion flow onto the primary SMBH is altered, leading to epochs of interrupted or excess accretion, manifesting as characteristic deep
dips and recoveries in the decline light curve \citep{Liu2009, Ricarte2016}.  
  
Initially, TDEs were thought to be single events and repeat flaring from the same source would falsify a TDE interpretation. Meanwhile, it has been shown that rates are enhanced in stages of galaxy merger evolution up to one event per decade in extreme cases, allowing for repeat flaring from the same system \citep{Chen2009}.{\footnote{Another mechanism for repeated outbursts of TDEs is partial tidal disruptions in single or binary SMBH systems \citep[e.g.][]{Krolik2011, Ryu2022a}}} 

\paragraph{Recoiling SMBHs} 
Numerical relativity simulations of coalescing SMBH binaries predict a population of recoiling SMBHs with velocities up to thousands km/s \citep[e.g.][]{Lousto2011}. These SMBHs would
leave their host galaxies altogether or oscillate about the centers of their hosts. Off-nuclear TDEs are unique signposts of such recoiling SMBHs \citep{KomossaMerritt2008}.   

\paragraph{Accretion and outflow physics in the (super-)Eddington regime} 
The (super-)Eddington accretion regime is still poorly understood. TDEs uniquely trace this regime and on very short timescales. 
More than that; within years to a decade we see the events move through regimes from super-Eddington to sub-Eddington, to 10$^{-2...-3}$ Eddington when the accretion mode itself fundamentally changes again. 
% see CL-papers for instance
Regular AGN move through these regimes only within millions of years (with the exceptions of a few extreme changing-look AGN; Sect. 3) and many may never pass the highly super-Eddington regime at all, or they do so only in the early Universe during epochs of major SMBH growth not yet accessible through observations.  

\paragraph{}
X-ray observations, and XMM-Newton in particular, have started to tap into this vast potential.
A rich discovery space has opened up. 
Here, we will review the salient results that have been obtained in recent years, with emphasis on those that go beyond the early results already obtained in the ROSAT era.  
For a broader review, including non-X-ray properties of TDEs, see \citet{Komossa2015}.

\subsection{Regular TDEs, spectral and light curve evolution} 

The first XMM-Newton observation of a TDE \citep{Komossa2004} was a follow-up of RXJ1242--1119 (R.A.-2000: 12$^{\rm h}$ 42$^{\rm m}$ 38.55$^{\rm s}$ decl.-2000: --11$^{\rm o}$ 19$^{'}$ 20.8$^{''}$; $z=0.05$); the second-ever identified TDE with an observed peak X-ray luminosity of at least $10^{44}$ erg/s in an inactive galaxy (absence of any broad or narrow emission lines in optical spectra). RXJ1242--1119 was first detected and followed up with ROSAT and was characterized by an exceptionally soft X-ray spectrum with photon index $\Gamma_{\rm x}\simeq-5$ or $kT_{\rm BB}=0.06$ keV \citep{KomossaGreiner1999}. With XMM-Newton and Chandra observations in 2001, 9 yrs after the peak, it was confirmed that fading X-ray emission was still coming from the event. The late-state X-ray spectrum of the event measured with XMM-Newton showed a significant hardening ($\Gamma_{\rm x}=-2.5$). This characteristic spectral hardening of an initially super-soft spectrum,  first seen in NGC 5905 \citep{KomossaBade1999}, was later found to be a characteristic of the majority of similar TDEs detected with recent X-ray missions with late-time follow-ups, while other sources retained their super-soft spectra
%%Donato2014: EUVE source in Abel cluster, but they focus on the initially soft EUVE/Chandra spectrum. At late stage, no spectra.
%% E2008: evidence for spectral hardening with XMM, but soft at peak
%% Lin2011: no X follow-up, but soft at peak 
%% Lin2015: spectrum slightly softer at lower flux. 
%% Lin2018: off-nuclear, remained soft. 
%% M2010: no-ci
\citep[e.g.][]{Esquej2008, Saxton2012,  Maksym2013, Donato2014, Lin2015, Li2020, Sazonov2021} \citep{Lin2018, Saxton2017}. 

At the late stage of evolution of RXJ1242--1119, the X-ray flux droped below the early-phase decline law.
%% not included in (Fig. \ref{fig:ROSAT}). 
The total amplitude of decline in the X-ray luminosity of RXJ1242--1119 was a factor of $>$1500 measured during the
last Chandra observation \citep{Komossa2005}. 

\subsection{High-resolution absorption spectroscopy}

Most TDEs were only observed with CCD-type spectral resolution since they were not bright enough during X-ray (follow-up) observations for X-ray grating spectroscopy. The first RGS spectrum of a TDE was taken of the X-ray bright event ASASSN14li 
\citep{Miller2015} at redshift $z=0.02$. It revealed several narrow absorption lines from highly ionized transitions of argon and sulphur superposed on a very soft thermal continuum of $kT \sim 0.05$ keV. The lines' presence was independently confirmed with Chandra. Velocity shifts on the order of a few 100 km/s
were interpreted as a sign of outflow or rotation. 
The highly ionized material has a column density of order few $10^{21}$ cm$^{−2}$. 
The absorption lines are variable in their velocity shifts.
The variability was used to argue for an origin close to the SMBH, in form of either a rotating wind from an inner, nascent accretion disk, or a filament of the disrupted star itself \citep{Miller2015}. 

\subsection{Long-lived TDEs}

The majority of TDEs have shown a rapid decrease after peak, with a light curve that is consistent with the predicted $t^{-5/3}$ decline that is characteristic for solar-type star disruptions in the high-viscosity regime. 
A notable exception is the case of XMMJ150052.0+015452 at $z=0.145$ identified in the XMM-Newton X-ray source catalogue \citep{Lin2017}. Its bright phase has already lasted for 15 years. After a rapid rise within 4 months to a peak luminosity of 
$3 \times 10^{43}$ erg/s it has only shown a very mild decline by a factor of $\sim$3  \citep{Lin2022}.  Its X-ray spectra remained very soft at $kT \sim 0.02$ keV without significant spectral hardening at late times. Its host galaxy is a dwarf starburst galaxy with an HII-type optical spectrum. The observations have been interpreted as  
TDE with a long-lasting, initially super-Eddington accretion phase in a low-viscosity regime as predicted by simulations of \citet{GR2015}.   

\subsection{Binary SMBH detection via TDE light curves}

The TDE light curve of SDSSJ120136.02+300305.5, identified in the XMM-Newton slew survey, does not show a smooth decline, but rather exhibits epochs of rapid, deep fades and recoveries above the initial decline law \citep{Saxton2012}.{\footnote{No radio-jet emission was detected from that system.}}  
These are the characteristic signs of a TDE evolving in a binary SMBH system \citep{Liu2009}.  
Detailed simulations of this system have shown that its light curve is well reproduced with a binary SMBH model with an orbital time scale of 150 d, a primary SMBH of mass $10^7$ M$_{\rm \odot}$, an SMBH mass ratio of $q \sim 0.08$ and a separation of $r=0.6$ milli parsec \citep{Liu2014}. 
The life time of this tightly bound binary system due to the emission of gravitational waves is on the order of 10$^6$ yrs. 
The host galaxy of this event is inactive, and this method therefore provides us with a means of detecting binary SMBHs in quiescent systems, while all other current binary detection methods require a long-lived accretion disk, i.e. AGN activity.  

%% Ueberleitung
\subsection{The highest amplitudes of variability}

Besides the jetted, beamed TDE Swift J1644+57 \citep{Bloom2011, Burrows2011}, several of the TDEs in quiescent galaxies identified with ROSAT and XMM-Newton showed amplitudes of X-ray variability exceeding a factor 1000 \citep[RXJ1242-1119, NGC 5905, RXJ1624+7554, SDSSJ1201+3003;][]{Halpern2004, Komossa2005, Saxton2017}; much higher than classical AGN variability, and long holding the record for the highest amplitudes of X-ray variability recorded from the centers of galaxies.
%% {\footnote{Only recently, 
%%%some AGN at such high amplitudes of variability were identified [GSN069 (Miniutti2021--only factor 200) 
%%and 1ES 1927+654 (Ricci2020)-- factor almost 10.000], but the variability mechanism was still suggested to be linked to TDE in the case of GSN069 despite the presence of a long-lived accretion disk.}}   
At the same time, high-amplitude variability -- not only in X-rays but also the lower-wavelength optical regime  -- in changing-look AGN has motivated new theoretical studies of accretion-disk instabilities and other mechanisms of high-amplitude changes in systems that harbor long-lived accretion disks. The second part of this review addresses these AGN.   

\section{Changing-look AGN}

The term changing-look AGN was initially coined to describe systems with high-amplitude variability in their X-ray properties \citep{Matt2003}. 
More recently, however, it has been widely used to refer to spectroscopic AGN-type changes between type 1 and type 2 in the optical band \citep[e.g.][]{LaMassa2015}, and we use it in this context in the following. These systems also exhibit high-amplitude variability in their spectral energy distributions (SEDs). 

\subsection{Extreme cases}

\begin{figure}
\centering
\includegraphics[clip, trim=1.9cm 13.0cm 1.8cm 3.5cm, width=\columnwidth]{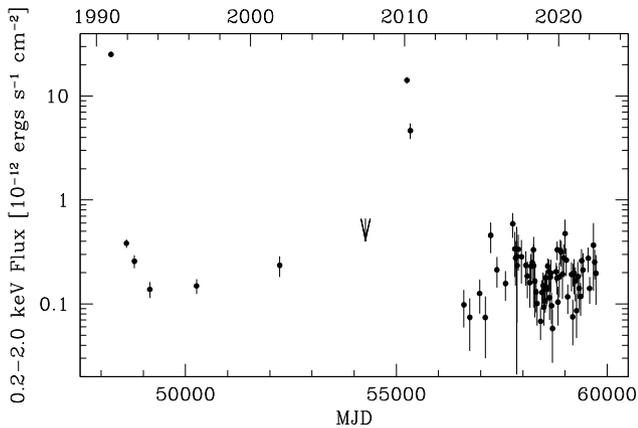}
    \caption{Longterm X-ray light curve of the changing-look AGN IC 3599 between 1990 and 2022 showing two bright X-ray outbursts with an amplitude of a factor $>$100 separated by 20 years.   
    }
    \label{fig:IC3599}
\end{figure}
 
One of the most extreme changing-look AGN is the Seyfert galaxy IC 3599 (REJ1237+264; $z=0.02$). It has shown two giant-amplitude X-ray outbursts (Fig. \ref{fig:IC3599}). The first one, in 1990, at a peak luminosity of $10^{44}$ erg/s 
%% 1.1 10{44} in 0.1-2.5 band, when NH fixed at Galactic
was detected with ROSAT \citep{Brandt1995, Grupe1995}, the 
second one, of similar luminosity and amplitude, 20 years later in 2000 with Swift \citep{Grupe2015}. Optical spectra taken during the first outburst revealed bright optical emission lines, including broad Balmer lines and high-ionization iron lines \citep{Brandt1995} that then faded away strongly in subsequent years \citep{Grupe1995, KomossaBade1999}. 
In low-state, IC 3599 is classified as Sy 1.9 galaxy with faint broad-line emission only detectable in H$\alpha$ and with a long-lived NLR including faint high-ionization iron lines \citep{KomossaBade1999}.

\begin{figure*}[ht]
\centering
\includegraphics[clip, trim=0.9cm 5.5cm 0.9cm 12.8cm, width=15.5cm]{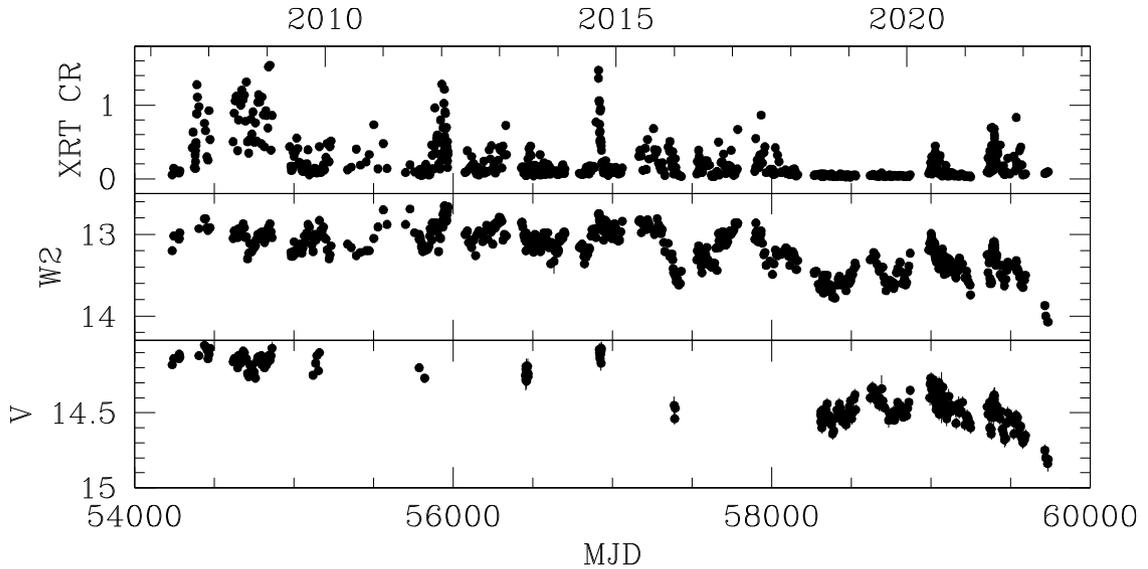}
    \caption{Swift longterm X-ray, UV (W2) and optical (V) light curve of Mrk 335 between 2008 and 2022. XMM-Newton follow-up spectroscopy was obtained at several of the extreme high- and low-states. Mrk 335 is highly variable. While the optical and UV variability are correlated with the UV leading by $\Delta t=1.5$ d \citep{Komossa2020}, the X-ray and UV variability are decoupled most of the time. During an epoch of dense HeII$\lambda$4686 and X-ray monitoring, HeII did not follow the ionizing X-ray continuum, making Mrk 335 a frozen-look AGN and implying that the BLR does not see the same X-ray continuum as the observer. 
    }
    \label{fig:M335}
\end{figure*}

The X-ray outbursts of the AGN IC 3599 were explained as thermal instability in the long-lived accretion disk, propagating through the inner region at the sound speed and causing an increase of the
local accretion rate. This causes episodes of repeat flaring on the timescale of decades when the inner disk is emptied and refilled \citep{Grupe2015}. 
Other scenarios to produce repeat outbursts,  like an OJ 287-type binary SMBH,  remain possibilities 
% (Komossa2014, Grupe2015) 
and will be tested in ongoing monitoring campaigns \citep{Grupe2022}. 

Other well-studied cases of remarkable long-term evolution of their SEDs 
accompanied by changes of their Seyfert type include 
Mrk 590 that decreased by a factor 100 over two decades in the optical band \citep{Denney2014},  
NGC 2617 that allowed to establish the X-rays as driver of the UV--NIR changes during a 70 day dense monitoring campaign \citep{Shappee2014}, NGC 1566 that changed its Seyfert type multiple times between 1970 and 2018 \citep{Alloin1986, Oknyansky2019, Parker2019}, and 1ES\,1927+654 that was characterized by particularly rapid and high-amplitude X-ray variability \citep{Trakhtenbrot2019, Ricci2020}.   

\subsection{Large samples and dedicated XMM-Newton searches}

Large optical spectroscopic data bases or dedicated optical spectroscopy of large samples of AGN have allowed an assessment of the frequency of type-changes among Seyfert galaxies and quasars \citep[e.g.][] {MacLeod2016, Runco2016, Ruan2016, Yang2018, Green2022, Wang2022}. 
Extreme type changes are found to be more common in Seyfert galaxies with 1\% of extreme changes between type 1 and type 2 \citep{Runco2016} in comparison to 0.1\% such changes in quasars \citep{MacLeod2016}. However, smaller type changes (between subtypes 1.0--1.5--1.8--2.0) are much more common in all samples.  
%% Wang2022 find CL-AGN more common in low-luminosity AGN.

The large optical archival samples are optimal for statistical studies. They lack simultaneous X-rays or rapid multi-wavelength (MWL) follow-ups.  
An alternative approach that was followed consists of  near-real time searches for X-ray or MWL flaring or dipping events for instance in XMM-Newton, Swift or Gaia data, that are then combined with rapid follow-up optical spectroscopy \citep{Schartel2007}. Several new changing-look AGN and repeat type-changers were identified that way \citep[e.g.][]{Parker2016, Komossa2017,  Kollatschny2018}, \citep{Parker2019}. Among them, IRAS 23226--3843, changing from type 1 to 1.9, stands out due to its extremely broad and double-peaked broad-line profile \citep{Kollatschny2020}.  

\subsection{Explanations}

The fact that emission-line fluxes from the BLR change along with the optical-UV (and X-ray) continuum emission in changing-look AGN implies that we see true changes in the ionizing continuum luminosity (as opposed to effects of absorption along our line-of-sight that could alter the observed X-rays).  
The question is then raised what drives these continuum changes. Given their high amplitude of variability, changes in accretion rate have generally been suspected as the underlying cause.  A challenge then is to explain the fastness of the variations of the order of decades to years. 
Much longer timescales would be expected for changes in the optical, if these propagated across the disk at the viscous timescale of order thousands of years \citep[e.g.][]{Lawrence2018}. 

The majority of models for changing-look AGN suggested in recent years have therefore explored different possibilities of temporarily enhancing accretion-rate changes, for instance through disk instabilities. 
Suggested explanations of changing-look AGN include the radiation-pressure disk instability, the Hydrogen ionization instability, magnetic pressure supported disks, large-radius instabilities in warped disks, or tidal interactions between two disks in binary SMBH systems \citep[e.g.][]{LightmanEardley1974, Nicastro2000, Grupe2015, Ross2018, NodaDone2018, DexterBegelman2019, WangBon2020, Igarashi2020, RajNixon2021, Sniegowska2022, Kaaz2022}.   

\subsection{Frozen-look AGN}

We introduce the term frozen-look AGN to describe systems that do not change their broad-line properties despite strong apparent changes in the {\it observed} ionizing continuum. 
% A bona fide example is also wpvs007

The nearby narrow-line Seyfert 1 (NLS1) galaxy Mrk 335 was once among the X-ray brightest AGN but has remained in
deep X-ray low-states most of the time in recent years, except
for occasional episodes of rapid flaring \citep[][our Fig. \ref{fig:M335}]{Grupe2007, Gallo2018, Komossa2020}. During a period of our dense monitoring of Mrk 335 with Swift, the AGN was also target of an optical reverberation mapping campaign \citep{Grier2012} for a duration of four months. Despite strong changes in X-ray emission at that epoch by a factor 5.4, the optical broad HeII$\lambda$4686 line remained nearly constant, even though its ionization potential is in the soft X-ray regime (54 eV), and therefore the line should be driven by the soft X-ray continuum. The lack of response of HeII to the {\it observed} X-ray variability implies that the BLR does not see the strong X-ray changes we observe along our line-of-sight \citep{Komossa2020}. Absorption affecting the observed X-rays can explain this observation, and can also explain the de-coupling between X-ray and UV variability. 
 
Because HeII is driven by the soft X-ray emission, simultaneous monitoring of X-rays and HeII or other high-ionization lines from the BLR is a powerful method of constraining the X-ray variability mechanism.  

%\begin{figure}[t]
%\centering
%\includegraphics[width=7.8cm, height=5.3cm]{fig5}
%    \caption{
%    }
%    \label{fig:alpha}
%\end{figure}

%\backmatter

\section*{Acknowledgments}
%% We would like to thank .... 
This research is partly based on observations obtained with XMM-Newton, an ESA science mission with instruments and contributions directly funded by ESA Member States and NASA. 
This research has made use of the
 XRT Data Analysis Software (XRTDAS) developed under the responsibility
of the ASI Science Data Center (SSDC), Italy.
This work made use of data supplied by the UK Swift Science Data Centre at the University of Leicester \citep{Evans2007}.
%% This research has made use of the NASA/IPAC Extragalactic Database (NED) which is operated by the Jet Propulsion Laboratory, California Institute of Technology, under contract with the National Aeronautics and Space Administration.

%\subsection*{Author contributions}

% \subsection*{Financial disclosure}

% None reported.

%\subsection*{Conflict of interest}
% The authors declare no potential conflict of interests.

%\bibliography{Wiley-ASNA}%

%\subsection{Bibliography}

\end{document}